\documentclass{appolb_modified}
\usepackage{graphicx}
\usepackage{bbm}
\usepackage{amsmath} 
\usepackage{amssymb}
\usepackage{amsthm}
\newcommand{\mnu}{\mathcal{M}_\nu}
\newcommand{\zz}{\mathbbm{Z}}
\newcommand{\bone}{\mathbbm{1}}
\newtheorem{theorem}{Theorem}
\newcommand{\diag}{\mbox{diag}}
\begin{document}
\title{{\small\hfill UWThPh-2013-30} \vspace{0.6cm} \\ 
Residual symmetries in the lepton mass matrices%
\thanks{Talk presented at XXXVII International Conference of
  Theoretical Physics, ``Matter to the Deepest,'' Ustro\'n, September
  1-6, 2013}%
}
\author{Walter Grimus
\address{University of Vienna, Faculty of Physics \\ Boltzmanngasse 5,
  A--1090 Vienna, Austria}
}
\maketitle
\begin{abstract}
It has been suggested that residual symmetries in the charged-lepton and
neutrino mass matrices can possibly reveal the flavour symmetry group
of the lepton sector. 
We review the basic ideas of this purely group-theoretical approach and
discuss some of its results. Finally, we also list its shortcomings.
\end{abstract}
\PACS{11.30.Hv, 14.60.Pq}
  
\section{Introduction}
Recent measurements of reactor neutrinos have
clearly demonstrated that the mixing angle $\theta_{13}$ of the lepton
mixing matrix $U = \left( U_{\alpha j} \right)$ 
($\alpha = e,\mu,\tau$, $j=1,2,3$) and, therefore, the element
$|U_{e3}| \equiv s_{13}$ is non-zero. 
Consequently, tri-bimaximal mixing (TBM)~\cite{HPS} is
not viable anymore.
Writing the mixing matrix in terms of its column vectors $u_j$ ($j=1,2,3$),
\textit{i.e.}
\begin{equation}\label{U}
U = (U_{\alpha j}) = \left( u_1, u_2, u_3 \right),
\end{equation}
we can say that $u_3$ cannot have the 
TBM form. However, it could still be
that $u_1$ or $u_2$ 
agrees with the corresponding column in TBM. These cases
are called TM$_1$ and TM$_2$, respectively, in~\cite{albright}:
\begin{equation}\label{TM12}
\mbox{{TM}}_1: \quad
u_1 = \frac{1}{\sqrt{6}} \left( \begin{array}{c} 2 \\ -1 \\ -1 
\end{array} \right),
\quad
\mbox{TM}_2: \quad
u_2 = \frac{1}{\sqrt{3}} \left( \begin{array}{c} 1 \\ 1 \\ 1 
\end{array} \right).
\end{equation}
Let us compare the predictions of these two cases.
In both cases, $s_{12}^2$
and the product $\cos\delta \tan 2\theta_{23}$, where $\delta$ is the
CKM-type phase in $U$, are determined by $s_{13}$. 
In the first case, these quantities are given by~\cite{albright}
\begin{equation}\label{TM1}
\mbox{TM}_1: \quad
s^2_{12} = 1 - \frac{2}{3 c_{13}^2} < \frac{1}{3}, \quad
\cos \delta \tan 2\theta_{23} \simeq 
-\frac{1}{2\sqrt{2}s_{13}} \left( 1 - \frac{7}{2} s_{13}^2 \right),
\end{equation}
whereas in the second case one obtains~\cite{albright,trimaximal}
\begin{equation}\label{TM2}
\mbox{TM}_2: \quad
s^2_{12} = \frac{1}{3 c_{13}^2} > \frac{1}{3}, \quad
\cos \delta \tan 2\theta_{23} \simeq 
\frac{1}{\sqrt{2}s_{13}} \left( 1 - \frac{5}{4} s_{13}^2 \right).
\end{equation}
In these formulas, we use the customary abbreviations $c_{ij}^2 \equiv
\cos^2 \theta_{ij}$ and $s_{ij}^2 \equiv \sin^2 \theta_{ij}$.
Moreover, 
in the formulas for $\cos \delta \tan 2\theta_{23}$, small terms of order
$s_{13}^4$ have been neglected for simplicity.
It should be noted that the best-fit values of $s_{12}^2$ are smaller than
1/3~\cite{forero}, but at any rate, 1/3 is in the $3\sigma$ range of
$s_{12}^2$. In that sense, TM$_1$ fits the data slightly better than
TM$_2$. Unfortunately, present data do not allow to determine the quantity 
$\cos \delta \tan 2\theta_{23}$.

Of course, it could also be that the ``true'' mixing matrix is not related to
tri-bimaximal mixing at all; in this case none of the columns of the
TBM matrix is a sensible approximation to one of the columns 
in the ``true'' mixing matrix. 

Recently, a purely group-theoretical attempt to track down the flavour group
$G$ of the lepton sector has been put forward which postulates that residual
symmetries in the charged-lepton and neutrino mass matrices are reflections of
$G$ and that the full flavour group can be determined by assembling the residual
symmetries from both mass matrices~\cite{lam,smirnov}.
This approach is based on the fact that at low energies the Standard
Model gives the correct gauge structure of any extension trying to explain
mass and mixing phenomena. Therefore, left-handed charged and
neutral lepton fields are not only in the same gauge doublets but also
in the same multiplets with respect to $G$, a fact which is hidden by
spontaneous symmetry breaking. 
In this approach, it is possible that the flavour group determines one
column of $U$ in terms of numbers of a purely group-theoretical origin.
It is also possible that two columns are determined, 
which means that, due to unitarity, the \emph{full} mixing matrix $U$ has
a group-theoretical origin. In this context, we will see that
TM$_2$ plays a role.

The plan of this report is as follows. 
In section~\ref{residual symmetries}, we discuss the idea of residual
symmetries. In section~\ref{group scans}, we briefly review
the results of~\cite{lindner}, based on a computer-algebraic group scan.
Then we point out the connection between residual symmetries and roots
of unity in section~\ref{residual symmetries and roots of unity}. After 
applying this in section~\ref{TM1 and roots of unity}
to find all flavour groups which enforce TM$_1$,
we conclude in section~\ref{caveats}
by discussing some caveats of the method of residual symmetries.

\section{Residual symmetries and lepton mixing}
\label{residual symmetries}
Though the method of residual symmetries can also be applied to Dirac
fermions, here we confine ourselves to Majorana
neutrinos. Then the mass Lagrangian in the lepton sector is given by
\begin{equation}
\mathcal{L}_\mathrm{mass} = 
-\bar \ell_L M_\ell \ell_R + 
\frac{1}{2} \nu_L^T C^{-1} \mathcal{M}_\nu \nu_L + \mbox{H.c.},
\end{equation}
where $C$ is the charge-conjugation matrix. The mass matrices of the
charged leptons and neutrinos are
diagonalized as
\begin{equation}\label{diagonalization}
U_\ell^\dagger M_\ell M_\ell^\dagger U_\ell = 
\mbox{diag} \left( m_e^2, m_\mu^2, m_\tau^2 \right)
\; \mbox{and} \;
U_\nu^T \mathcal{M}_\nu U_\nu = \mbox{diag} \left( m_1, m_2, m_3 \right),
\end{equation}
respectively, 
leading to the mixing matrix $U = U_\ell^\dagger U_\nu$. 
The fact that these matrices are diagonalizable can be reformulated as 
\begin{equation}\label{invariance}
V_\ell(\alpha)^\dagger M_\ell M_\ell^\dagger V_\ell(\alpha) = 
M_\ell M_\ell^\dagger, \quad
V_\nu(\epsilon)^T \mathcal{M}_\nu V_\nu(\epsilon) = \mathcal{M}_\nu,
\end{equation}
with unitary matrices $V_\ell(\alpha)$, $V_\nu(\epsilon)$ defined as
\begin{eqnarray}
V_\ell(\alpha) &=&
U_\ell \,\mbox{diag} \left( e^{i\alpha_1}, e^{i\alpha_2}, e^{i\alpha_3} \right) 
U_\ell^\dagger,
\\ 
V_\nu(\epsilon) &=&
U_\nu \,\mbox{diag} \left( \epsilon_1,  \epsilon_2,  \epsilon_3 
\right) U_\nu^\dagger.
\end{eqnarray}
Equation~(\ref{invariance}) holds for arbitrary $\alpha_j$ and 
arbitrary $\epsilon_j = \pm 1$.
In group-theoretical terms, the mass matrices are invariant under
\begin{equation}
V_\ell(\alpha) \in U(1) \times U(1) \times U(1),
\quad
V_\nu(\epsilon) \in \mathbbm{Z}_2 \times \mathbbm{Z}_2 \times
\mathbbm{Z}_2.
\end{equation}
Obviously, 
$V_\ell(\alpha)$ and $V_\nu(\epsilon)$ depend on the vacuum expectation
values and Yukawa coupling constants, and
equation~(\ref{invariance}) contains no information beyond
diagonalizability.

The idea of residual symmetries is the following.
In a weak basis, the fields $\ell_L$, $\nu_L$ are in the same
multiplet of the flavour group $G$ 
under which the Lagrangian is invariant. The flavour group
$G$ is broken to different subgroups $G_\ell$ and $G_\nu$ in the charged-lepton 
and neutrino sector, respectively. From equation~(\ref{invariance}) we
know that
\begin{equation}\label{GG}
G_\ell \subseteq U(1) \times U(1) \times U(1), \quad
G_\nu  \subseteq \mathbbm{Z}_2 \times \mathbbm{Z}_2 \times \mathbbm{Z}_2.
\end{equation}
For simplicity we assume that there is 
one generator $T$ of $G_\ell$ and one generator $S$ of
$G_\nu$. Therefore, we have 
\begin{equation}
T^\dagger M_\ell M_\ell^\dagger T = M_\ell M_\ell^\dagger, \quad
S^T \mnu S = \mnu.
\end{equation}
We furthermore assume that $T$ has three different eigenvalues. 
Then $T$ and $S$ determine one column of $U$, as we will argue now.

Due to equation~(\ref{GG}), we have
\begin{equation}
S^2 = \bone \quad \Rightarrow \quad S = \pm (2 uu^\dagger  - \bone) 
\end{equation}
with a unit vector $u$ and $Su = \pm u$. 
Since $T$ commutes with $M_\ell M_\ell^\dagger$ and has three
different eigenvalues, we know that
$U_\ell^\dagger T U_\ell = \widetilde T$ is diagonal.
Therefore, $U_\ell$ is determined by $T$ and is thus independent of
any parameters of the Lagrangian.
For the rest of the argument we use the following theorem.
\begin{theorem}
If $S^T \mnu S = \mnu$ with $S = \pm (2uu^\dagger - \bone)$, 
then $\mnu u \propto u^*$.
\end{theorem}
\noindent
Thus, $u$ is, apart from a phase, one of the columns of $U_\nu$ and,
therefore, $U_\ell^\dagger u$ is a column in the mixing matrix $U$.
Because $U_\ell^\dagger u$ is determined by the group, it does not contain
parameters of the model.

If there are two matrices $S_1$, $S_2$ with $S_j^T \mnu S_j = \mnu$
and $S_1 S_2 = S_2 S_1$, two columns of $U$ are determined and thus
the complete mixing matrix.

There are two ways to tackle the mathematical problem of residual
symmetries for the determination of possible flavour symmetry groups:
\begin{enumerate}
\item
Scanning classes of finite groups,
\item
solving relations involving roots of unity.
\end{enumerate}

\section{Group scans}
\label{group scans}
Scans of groups have for instance been performed in~\cite{lindner,lam1} using
GAP~\cite{GAP} and the small groups library~\cite{SGL}. This library
contains all finite groups with order up to 2000, with the
exception of the order of 1024. Here we want to discuss the results
of~\cite{lindner}. The authors of this paper have assumed that
$G_\nu = \zz_2 \times \zz_2$, \textit{i.e.}\ there are two matrices
$S_j$ in $G_\nu$, and that the group produces mixing parameters
$s^2_{ij}$ within the $3\sigma$ range of the fit results
of~\cite{forero}. The authors have performed two scans. In the first
one, they allowed for $\mbox{ord}\, G < 1536$, with the exception of 
one group whose order is just 1536, and assumed that $G_\ell$
is generated by
$\widetilde T = \diag (1 ,\omega, \omega^2 )$ with $\omega = e^{2\pi i/3}$.
It is amazing that only three groups, namely
$\Delta(6 \times 10^2)$,
$(\zz_{18} \times \zz_6) \rtimes S_3$ 
and
$\Delta(6 \times 16^2)$,
lead to acceptable mixing patterns. 
All three groups have TM$_2$ and a trivial CKM-type phase. 
Therefore, $s_{12}^2$ is given by
equation~(\ref{TM2}). One can show~\cite{GL2013} that, in the case of
the three viable groups, for every
$s_{13}^2$ there are two solutions of $s_{23}^2$ given by
\begin{equation}
s_{23}^2 = \frac{1}{2} \left( 1 \pm 
\frac{\sqrt{2 s_{13}^2 - 3 s_{13}^4}}{c_{13}^2} \right).
\end{equation}
With this formula, the numbers in the third colum in table~3
of~\cite{lindner} are reproduced.

In the second scan the authors of~\cite{lindner} have assumed 
that the group order is smaller that 512 but allowed for $G_\ell$ 
\emph{all} Abelian finite groups. In this case, no candidates were found.

\section{Residual symmetries and roots of unity}
\label{residual symmetries and roots of unity}
Now we come to the second way of treating residual
symmetries. Let us assume for simplicity that 
\begin{itemize}
\item
$G_\ell$ is generated by $T$ and $G_\nu$ by $S$, respectively,
\item
$\det S = 1$ and thus $S = 2uu^\dagger - \bone$.
\end{itemize}
Then finiteness of the group $G$ requires the 
existence of positive integers $m$, $n$ such that 
\begin{equation}
T^m = S^2 = (ST)^n = \bone.
\end{equation}
We denote the eigenvalues of $T$ by $e^{i\phi_\alpha}$. If $ST$ has eigenvalues
$\lambda_j$, then $\lambda_j^n = 1$. If the unit vector $u$ in $S$
coincides with the $i$-th column of $U = (U_{\alpha j})$, then the 
trace and determinant of $ST$ give~\cite{smirnov}
\begin{equation}\label{6roots}
\sum_{\alpha = e,\mu,\tau} \left( 2\left| U_{\alpha i} \right|^2 - 1
\right) e^{i\phi_\alpha} = \lambda_1 + \lambda_2 + \lambda_3
\quad \mbox{and} \quad
\prod_\alpha e^{i\phi_\alpha} = \lambda_1 \lambda_2 \lambda_3,
\end{equation}
respectively. Thus we have a vanishing sum of six roots of unity plus
the condition for the determinant.
Equation~(\ref{6roots}) can be used in two ways. 
Departing from elements $S$ and
$T$ of a known group $G$, one can search for suitable 
$\left| U_{\alpha i} \right|^2$ ($i=1,2,3$). 
One the other hand, for a given column in $U$, one can try
to find a suitable group $G$.

\section{TM$_1$ and roots of unity}
\label{TM1 and roots of unity}
As an application of equation~(\ref{6roots}) we want to investigate
which groups can enforce TM$_1$~\cite{grimus}.
In this case, the coefficients in equation~(\ref{6roots}) are
$2\left| U_{e 1} \right|^2 - 1 = 1/3$ and
$2\left| U_{\mu 1} \right|^2 - 1 = 
2\left| U_{\tau 1} \right|^2 - 1 = -2/3$. Therefore,
equation~(\ref{6roots}) leads to the vanishing sum
\begin{equation}\label{sumofroots}
-e^{i \phi_e} + 2 e^{i\phi_\mu} + 2 e^{i\phi_\tau} + 3 \lambda_1 +  
3 \lambda_2 +  3 \lambda_3 = 0.
\end{equation}
In order to solve this equation, one can apply a theorem of Conway and
Jones~\cite{conway}, referring to all possible vanishing sums of roots
of unity up to nine roots.\footnote{Actually, theorem~1
  in~\cite{grimus}, quoted from the book of Miller, Blichfeldt and
  Dickson, is wrong and one has to use the theorem of Conway and Jones
  instead, in order to find the solution of equation~(\ref{sumofroots}). 
  We thank R.~Fonseca for pointing this out us.}
Amazingly, the solution of equation~(\ref{sumofroots}) allows very
little freedom:
\begin{equation}\label{sol}
e^{i \phi_e} = \eta, \quad
e^{i\phi_\mu} = \eta \omega,  \quad
e^{i\phi_\tau} =\eta \omega^2,  \quad
\lambda_1 = \epsilon, \quad
\lambda_2 = -\epsilon, \quad
\lambda_3 = \eta,
\end{equation}
where $\eta$ is an arbitrary root of unity and $\epsilon = \pm i\eta$.
In the basis where the charged-lepton mass matrix is diagonal, which we
indicate by a tilde on $T$ and $S$, this solution leads to
\begin{equation}\label{tilde}
\widetilde T = \eta\, \diag \left( 1, \omega, \omega^2 \right), 
\quad
\widetilde S = 2 u_1 u_1^\dagger - \bone
\end{equation}
with $u_1$ of equation~(\ref{TM12}).
Clearly, $\widetilde S$ must have this form because we departed from
TM$_1$. One can show that $\widetilde T$ and $\widetilde S$ 
generate the group $\zz_q \times S_4$
with $\eta$ being a primitive root of order $q$~\cite{grimus}.
Thus with the method of residual symmetries one finds an almost unique
group for TM$_1$. For the generators of $S_4$ in its three-dimensional
irreducible representations see \textit{e.g.}~\cite{GL-review}.

It is instructive to go into another basis where $S$ has a
simpler form. Performing a similarity transformation with the matrix
$U_\omega$ defined for instance in~\cite{grimus},
we obtain
\begin{equation}
S = U_\omega \tilde S U_\omega^\dagger = 
\left( \begin{array}{rcc}
-1 & 0 & 0 \\ 0 & 0 & 1 \\ 0 & 1 & 0 
\end{array} \right),
\quad 
T = U_\omega \tilde T U_\omega^\dagger = 
\eta \left( \begin{array}{rrr}
0 & 1 & 0 \\ 0 & 0 & 1 \\ 1 & 0 & 0 
\end{array} \right).
\end{equation}
In that basis, $M_\ell M_\ell^\dagger$ is invariant under cyclic
permutations and the eigenvector $u$ of $S$ with eigenvalue~1 is given
by $u = (0,1,1)^T/\sqrt{2}$. Consequently, up to a basis transformation, 
the mechanism for TM$_1$ developed here boils down to
$U_\omega^\dagger u = u_1$, where $u_1$ is 
the first column of TBM~\cite{grimus, lavoura}.

\section{Residual symmetries and caveats}
\label{caveats}
Up to now we have operated under the premises that 
the residual symmetries in $M_\ell M_\ell^\dagger$ and $\mnu$ really
determine the flavour group $G$ as a symmetry group of the Lagrangian.
Let us be more precise now and denote by $\bar G$ the group generated by the
residual symmetries.
What is the possible relationship between $G$ and $\bar G$? 
Clearly, $\bar G \subset U(3)$ due to three families of fermions.
Since the method is purely group-theoretical and uses only information 
contained in the mass matrices, $\bar G$ can at most yield $D(G)$, the
representation of $G$ on the lepton gauge doublets. 
Moreover, there are models with accidental symmetries in the mass
matrices, in which case $\bar G$ not even a subgroup of $D(G)$.
Finally, there are predictive models with total breaking of
$G$. Clearly, there the method of residual symmetries is not applicable.
\\[1mm]
\noindent
\textbf{Acknowledgement:} The author thanks the organizers for their
hospitality and the pleasurable atmosphere at the conference.

\end{document}